\documentclass[aps,pra,twocolumn,groupedaddress,showpacs]{revtex4}

\usepackage{graphicx}%
\usepackage{dcolumn}
\usepackage{amsmath}
\begin{document}

\title{Measurement of a Mixed Spin Channel Feshbach Resonance in Rubidium 87}
\author{M.~Erhard}
\author{H.~Schmaljohann}%
\author{J.~Kronj\"ager}%
\author{K.~Bongs}%
\author{K.~Sengstock}%
\affiliation{%
Institut f\"ur Laser-Physik,
Luruper Chaussee 149,
22761 Hamburg,
Germany}

\date{\today}

\begin{abstract}
We report on the observation of a mixed spin channel Feshbach
resonance at the low magnetic field value of (9.09$\pm$0.01)\,G
for a mixture of $|2,-1\rangle$ and $|1,+1\rangle$ states in
$^{87}$Rb. This mixture is important for applications of
multi-component BECs of $^{87}$Rb, e.g. in spin mixture physics
and for quantum entanglement. Values for position, height and
width of the resonance are reported and compared to a recent
theoretical calculation of this resonance.
\end{abstract}


\maketitle

Feshbach resonances are a versatile tool to alter the scattering
properties of atomic ensembles in a controlled way, which opens
fascinating possibilities for Bose-Einstein condensates (BEC) as
well as ultra-cold Fermi gases. In the vicinity of a Feshbach
resonance the atom-atom interaction characterized by the s-wave
scattering length can typically be tuned over a wide range of
negative and positive values by simply varying an applied magnetic
field. Feshbach resonances have been observed for various alkali
atoms
\cite{Inouye1998a,Courteille1998a,Roberts1998a,Vuletic1999a,Cornish2000a,Loftus2002a,Khaykovich2002a,Strecker2002a,Dieckmann2002a,Marte2002a,O'Hara2002a,Bourdel2003a}.
The tunability of the interatomic interactions has been used to
alter the mean-field energy of a BEC leading to a collapse of the
condensate \cite{Donley2001a}. Furthermore the possibility of
coherently coupling atomic and molecular states has sparked recent
interest in atomic Feshbach resonances resulting in a series
of cold molecule experiments based on BEC and cold Fermi gases
\cite{Donley2002a,Regal2003c,Herbig2003a,Durr2003a,Cubizolles2003a,Jochim2003a,Strecker2003a,Regal2003d}.
For $^{87}$Rb, the element most commonly used in Bose-Einstein
condensation experiments, no Feshbach resonances have been found
for the magnetically trappable states $|F,m_F\rangle =
|1,-1\rangle$ and $|F,m_F\rangle = |2,+2\rangle$. In a recent
precision experiment on the $|1,+1\rangle$ state and a mixture of
the $|1,+1\rangle$ and $|1,0\rangle$ states more than 40 Feshbach
resonances have been observed~\cite{Marte2002a}. Most of them are
in excellent agreement with theory, which makes $^{87}$Rb one of
the elements with the most precisely known collisional parameters.
The observed Feshbach resonances are mostly
relatively narrow and at high magnetic field values of several
100\,G, making their exploitation a difficult task.

In this paper we report on the observation of an easily accessible
mixed spin channel Feshbach resonance between the $|2,-1\rangle$
and $|1,+1\rangle$ states of $^{87}$Rb at a low magnetic field
value. This resonance has been predicted theoretically based on
recent experimental data by E. G. M. van Kempen et
al.~\cite{Kempen2002a}.

The knowledge of atom-atom interaction parameters between
different spin states is fundamental for a deeper understanding of
so called spinor condensates. Lifetimes of spin mixtures in the
$F=2$ manifold as well as collective spin dynamics leading to
nanomagnetic effects are governed by the atom-atom interactions. A
tunability of spin-interactions in dilute quantum gases may
e.g.~improve the experimental feasibility of spin-squeezing
scenarios \cite{Soerensen2001a} leading to future applications in
quantum optics and quantum computation.

Our experimental apparatus is based on a double-MOT
(magneto-optical trap) system which can produce magnetically
trapped Bose-Einstein condensates of 10$^6$ atoms in the
$|2,+2\rangle$ state every 30..45\,s. These are subsequently
transferred into a far detuned optical dipole trap operated at
1064\,nm with trapping frequencies of approx.~$2\pi\times 890$\,Hz
vertically, $2\pi\times 160$\,Hz horizontally and $2\pi\times
20$\,Hz along the beam direction. The experiments reported were
performed typically with initially $10^5$ optically trapped atoms.
The confining potential is independent of the spin- and
hyperfine-state and is therefore well suited for examinations of
arbitrary spin- and hyperfine-states and mixtures of those
\cite{Schmaljohann2003a}. For detection, the atoms are released
from the dipole trap and separated by a Stern-Gerlach gradient of
$\approx 26$\,G/cm at an offset field of $\approx 157$\,G applied
for 7.5\,ms. After a further time of flight of typically 7\,ms an
absorption image is taken. The linear Zeeman effect leads to a
separation of 650\,$\mu $m between $m_F$-states. An additional
separation of 85\,$\mu m$ between $F$-states occurs due to the
quadratic Zeeman effect. Therefore each absorption image provides
population numbers for each of the $m_F$ components of the $F=2$
and $F=1$ states separately and simultaneously.

For the experiment reported here, a mixture of the $|2,-1\rangle$
and $|1,+1\rangle$ states is prepared. We use a Landau-Zener
crossing technique \cite{Mewes1997a} to transfer populations
between the $m_F$-states. An offset field of 25\,G during all
Landau-Zener processes leads to a significant difference of the
$m_F$-transition frequencies due to the quadratic Zeeman effect
and therefore allows specific addressing of the individual
transitions. Slowly sweeping the radio-frequency allows for an
adiabatic following of the eigenstate.

\begin{figure}[htbp]
  \begin{center}
    \includegraphics{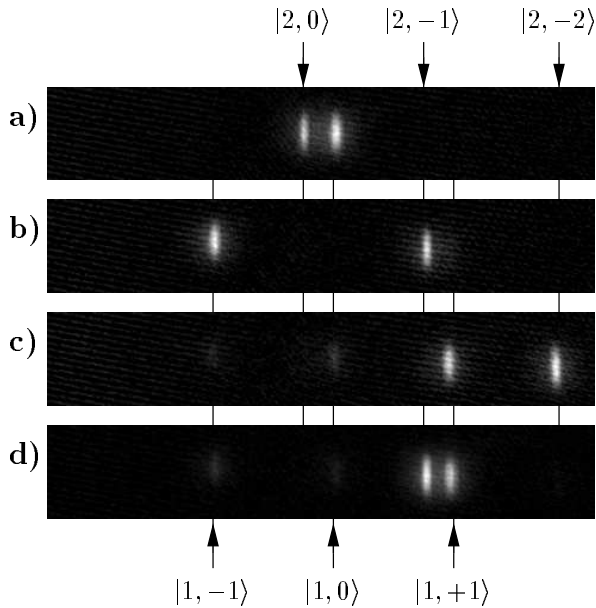}
    \caption{Absorption images after different preparation steps. The corresponding
    sweeps and states are summarized in table \ref{tab:1}.}
    \label{fig:1}
  \end{center}
\end{figure}

In order to prepare the desired spin mixture, we first transfer
the atoms initially in the $|2,+2\rangle$ state adiabatically into
the $|2,0\rangle$ state.
Subsequently the magnetic field is lowered to 10\,G and a
$\pi/2$-Raman pulse is used to transfer 50\% of the population
into the $|1,0\rangle$ state resulting in the distribution shown
in Fig.~\ref{fig:1}a.

The Raman laser system consists of two phase-locked diode lasers
similar to \cite{Prevedelli1995a}. The master laser is operated in
a free-running mode approximately 16\,GHz above the
$F=2\leftrightarrow F'=3$-transition resonance. The slave laser is
phase-locked to the master at a difference frequency of 6.8\,GHz
above the master laser frequency taking into account the quadratic
Zeeman shifts of 47\,kHz between the $|2,0\rangle$ and
$|1,0\rangle$ states. The intensities of the two equally and
circularly polarized Raman laser beams at the position of the
condensate are on the order of 30\,mW/cm$^2$ each and the pulse
duration is 100\,$\mu$s.
The mixture of the $|2,0\rangle$ and $|1,0\rangle$ states obtained
after the Raman pulse is transferred to the final mixture by 3
further Landau-Zener sweeps. Table \ref{tab:1} summarizes these
steps and corresponding absorption images are shown in
Fig.~\ref{fig:1}.
We would like to note at that point that the absence of linear
Zeeman shifts during the Raman passage made us favour the
implemented scheme in comparison to a simpler sequence (e.g. using
a Landau-Zener passage to $|2,-1\rangle$ and a $\pi/2$-pulse of
appropriately polarized Raman lasers to transfer half of the
population directly to the $|1,+1\rangle$). This way we achieve a
good reproducibility of the preparation.
\begin{table}
 \caption{Steps for the preparation of the $|2,-1\rangle$, $|1,+1\rangle$
 mixture, starting with a $|2,2\rangle$ sample in the optical dipole trap.} \label{tab:1}
 \begin{ruledtabular}
  \begin{tabular}{ccccc}
     Action && \multicolumn{2}{c}{Mixture}& Picture\\
            && State 1 & State 2 & \\
     \colrule
            && $|2,+2\rangle$ & & \\
     Sweep 1 &$\rightarrow$& $|2, 0\rangle$ & & \\
     Raman $\pi/2$ &$\rightarrow$ & $|2, 0\rangle$ & $|1, 0\rangle$ &a) \\
     Sweep 2 &$\rightarrow$& $|2, -1\rangle$ & $|1, -1\rangle$ &b)\\
     Sweep 3 &$\rightarrow$& $|2, -2\rangle$ & $|1, +1\rangle$ &c)\\
     Sweep 4 &$\rightarrow$& $|2, -1\rangle$ & $|1, +1\rangle$ &d)\\
  \end{tabular}
\end{ruledtabular}
\end{table}

In order to observe the Feshbach resonance, the prepared mixture
of $|2,-1\rangle$ and $|1,+1\rangle$ is held for a variable hold
time in the dipole trap while a precise current is applied to
Helmholtz coils inducing magnetic fields up to 10\,G. Then the
dipole trap is switched off and after Stern-Gerlach separation an
absorption image is taken. The number of atoms in the condensate
fractions for the $|2,-1\rangle$ and $|1,+1\rangle$ states are
determined by performing a 1d fit to the column sums of the
processed absorption images. For every value of the magnetic field
the number of atoms in the condensates for both states is
determined for negligible hold time as $N_1(0)$ and $N_2(0)$ and
subsequently for hold time $t_0$ as $N_1(t_0)$ and $N_2(t_0)$.

\begin{figure}[htbp]
  \begin{center}
    \includegraphics[width=86mm]{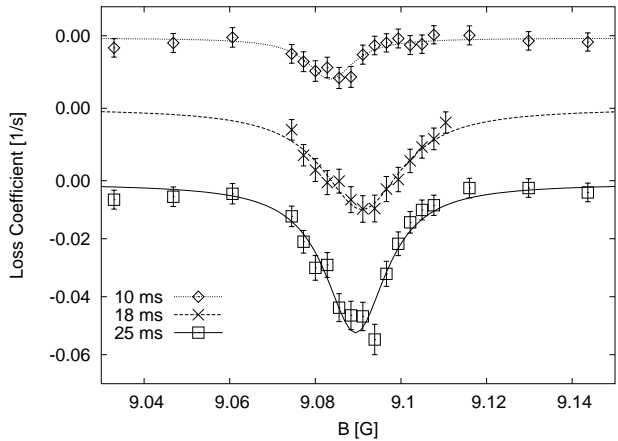}
    \caption{Loss coefficients for a mixture of $|2,-1\rangle$
    and $|1,+1\rangle$ states as function of the magnetic field for hold times of 10, 18 and
    25\,ms. The loss coefficient is proportional to the two-body loss rate $\bar{\gamma}$
    multiplied by the hold time (see text for further explanation).}
    \label{fig:2}
  \end{center}
\end{figure}

Note that for the Feshbach resonance investigated in this paper
the atoms in the incoming channels differ not only in their $m_F$-
but also in their $F$ quantum number leading to a significant
extension of the number of outgoing and loss channels as compared
to single spin channel resonances. For all channels not conserving
the hyperfine state or total spin the released hyperfine or Zeeman
energy leads to an instantaneous loss of atoms from the trap.
In the following we analyze the loss dynamics in order to determine position
and width of the resonance in a well defined way.
Loss during the hold time is evaluated assuming the following
differential equation which describes the particle number $N(t)$
in a harmonic trap as a function of time $t$ in presence of a
two-particle loss process \cite{Soding1999a}
\begin{equation}
  \dot{N} = \gamma(B(t)) N^{7/5}.
  \label{eq:eqm1}
\end{equation}
The loss rate, $\gamma(B(t))$ depends on the s-wave scattering
length, introducing a magnetic field dependence in order to allow
for temporally varying values (as the magnetic field
root-mean-square-noise is comparable to the resonance width). It
is important to annotate at this point that the equation above
assumes an adiabatic following of the trapping volume during the
decay process and therefore is not strictly valid for our
considered process due to the fact that the decay is fast compared
to the axial trapping frequency. Nevertheless the equation is a
reasonable approximation \cite{Erhard2003a} and allows the
introduction of a loss coefficient, $C$, characterizing particle
losses until time $t_0$. Variable separation of eq.~\ref{eq:eqm1}
yields \begin{equation}
  C = \bar{\gamma }(B) t_0 := \int_0^{t_0} dt \gamma(B(t)) =
  \frac{5}{2}\left(\frac{1}{N(0)^{2/5}}-\frac{1}{N(t_0)^{2/5}}\right),
\end{equation}
defining a time averaged loss rate $\bar{\gamma }$. The loss
coefficient is determined from the experiment as
\begin{equation}
  C = \frac{5}{2}\left(\frac{1}{(N_1(0)+N_2(0))^{2/5}}-\frac{1}{(N_1(t_0)+N_2(t_0))^{2/5}}\right)\quad\mbox{.}
\end{equation}
This equation can be applied due to $N_1 \approx N_2$ \cite{Erhard2003b}.
Figure \ref{fig:2} shows the according curves for hold times of
10, 18 and 25\,ms.
The data has been fitted by a Lorentzian function
\begin{equation}\label{eq:laplace}
  C(B) = C_0 + \frac{A}{1+4((B-B_0)/\Delta B)^2}\quad\mbox{.}
\end{equation}
extracting the parameters shown in table \ref{tab:2}. The $C_0$ value turns out to be small compared to the resonance depth and is consistent with two-body loss rates of other experiments \cite{Schmaljohann2003a}.
\begin{table}
 \caption{Fitting parameters for the experimental data shown in figure \ref{fig:2} using eq.~(\ref{eq:laplace}) and calculated loss rates $\bar{\gamma}=A/(t_0-6\mbox{\,ms})$ taking into account the initial ringing of the magnetic field (see text).}
\label{tab:2}
 \begin{ruledtabular}
  \begin{tabular}{ccccc}
     Hold time $t_0$ [ms] & $\bar{\gamma}(B_0)$ [1/s] & $B_0$ [G] & $\Delta B$ [G]\\
  \colrule
     10 & -3.5 & 9.084 & 0.013  \\
     18 & -2.8 & 9.091 & 0.023 \\
     25 & -2.7 & 9.089 & 0.017
  \end{tabular}
\end{ruledtabular}
\end{table}

The magnetic field is calibrated by performing Landau-Zener sweeps
within the $F=2$ manifold at the approximate magnetic field of the Feshbach
resonance.
\begin{figure}[htbp]
  \begin{center}
    \includegraphics[width=86mm]{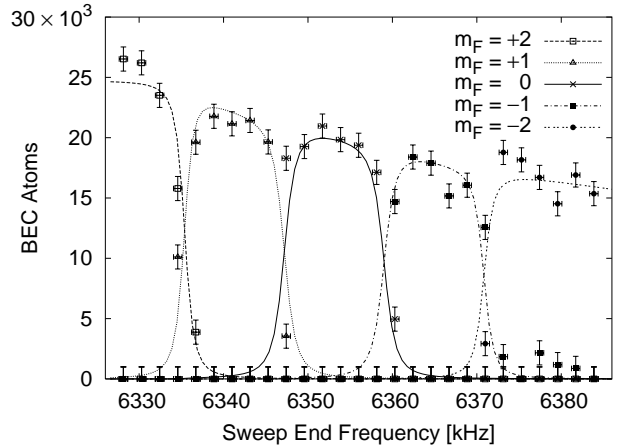}
    \caption{Final population of $m_F$ states after a Landau-Zener sweep starting at 6326\,kHz versus sweep end frequency. The horizontal errors represent the accuracy of the used sweeping generator. A theoretical model is fitted to the data to determine a magnetic field calibration yielding a linear Zeeman splitting frequency of $2\pi\times(6353.13\pm0.08)$\,kHz.}
    \label{fig:3}
  \end{center}
\end{figure}
Figure \ref{fig:3} shows measured atom numbers in the BEC fraction
for different end frequencies of the Landau-Zener sweep
starting at 6326\,kHz. The data
is compared to a theoretical model of the $m_F$ populations taking into
account a $m_F$-dependent particle loss and Landau-Zener parameters during
the sweeps.
The positions of the $m_F$ transition frequencies are evaluated by
a simultaneous fit of the theoretically calculated populations
for each of the $m_F$-components. Due to symmetry this calibration method is
first order insensitive to AC-Stark shifts connected to the coupling field.
The conversion to magnetic field values
is based on a Land\'e-factor of $g_F=0.49945$ for $^{87}$Rb. We want to mention
that the difference between measured resonance positions and the calibration
field of 9.088\,G leads to relative errors of the order of $10^{-5}$
taking into account our offset field compensation.
A detailed error budget estimating higher order terms of the Breit-Rabi formula and AC Stark shifts leads to an overall calibration error of $< 1$\,mG.

Nevertheless the observed width is
likely to be significantly broadened by current noise of the power
supply ($\hat{=}$5\,mG$_{rms}$) and AC stray magnetic fields in the
laboratory (on the order of 5\,mG$_{rms}$ in the vicinity of the
trapped atoms). The observed width of the Feshbach resonance is
thus consistent with the theoretically predicted value of 1-2\,mG
\cite{Kempen2002a,Kempen2003a}, while we find a slight shift of its offset on
the order of 3$\times 10^{-3}$, i.e.~30\,mG (theoretical value: 9.12\,G \cite{Kempen2002a}). Note that the initial ringing of the current
in the Helmholtz coils when switched on at the beginning of the
hold time inhibits the observability of the resonance for short
hold times $< 6$\,ms and leads to slight shifts of the resonance
mainly for short hold times (as observed for $t_0$=10\,ms,
compare fig.~\ref{fig:2} and table.~\ref{tab:2}).
Numerical integration of the
current-switching curve however yields a shift of the resonance of
less than 2\,mG for $t_0\geq 18$\,ms, which thus cannot account for the shift
we observe versus the theoretical prediction.
Nevertheless additional eddy currents may be present. This conclusively explains the shift of the observed resonance for $t_0=10$\,ms. Concerning longer hold times we observe no shift between the resonance curves for $t_0=18$\,ms and $t_0=25$\,ms and therefore shifts due to magnetic field switching and eddy currents seem to be unlikely for these hold times.

A major difference is found concerning the loss rate
$\bar{\gamma }\approx -2.8/ $s,
which is nearly two orders of magnitude lower than predicted \cite{erhard2003c}.
This can be explained in part by broadening of the resonance due to technical
noise. In addition, our estimation is based on a homogeneous mixture, but spacial separation effects of the two immiscible spin-components may reduce the overlap and lead to a lower loss rate than expected \cite{Kempen2003a}.

In conclusion we have measured a mixed spin channel Feshbach
resonance in $^{87}$Rb between the states $|2,-1\rangle$ and
$|1,+1\rangle$ at an easily accessible magnetic field of
9.09$\pm$0.01\,G. The line width is consistent with theoretical
predictions~\cite{Kempen2002a,Kempen2003a}, but there remain a
slight line shift of $\approx 30\, $mG and a discrepancy in loss
rates to be resolved.

{\it Note added.} Recently, observation of this resonance using
entanglement interferometry with pairs of atoms in an optical
lattice has been reported \cite{Widera2003a}.

We thank E.G.M.~van Kempen and B.J.~Verhaar for stimulating discussions and
acknowledge support from the Deutsche Forschungsgemeinschaft in the SPP 1116.


\begin{thebibliography}{32}
\bibitem[{Erh({\natexlab{a}})}]{Erhard2003a}
\bibinfo{note}{The equation describing a two-body loss reads $\dot{N}=\gamma
  N^\alpha$ with $\alpha=7/5$ for $\gamma\ll \omega_i$, $i=1,2,3$ ($\omega_i$
  are the trapping frequencies) and approximate $\alpha=2$ for $\gamma\gg \omega_i$ which
  may be considered as 'instantaneous decay'. Our experimental situation lies
  in between and the equation of motion can not be written in such a simple
  form. Evaluation of the experimental data assuming the $\alpha=2$ case
  (instead of $\alpha=7/5$) leads to a modification of 20\,\% in comparison of
  the loss rate to the theoretical result. The dynamics of the decay process is
  a interesting problem on its own and would go beyond the scope of this
  paper.}
\bibitem[{Erh({\natexlab{b}})}]{Erhard2003b}
\bibinfo{note}{The used preparation scheme fulfills the requirement $N_1\approx
  N_2$ with an accuracy of 20\,\%. The preparation asymmetry can be determined
  and compensated for each measurement regarding the actual asymmetry and
  assuming losses by pairs of $F=2$ and $F=1$ atoms. Nethertheless this
  procedure leads only to changes of 20\,\% in initial loss rates and has negligible
  effects on position and width of the resonance.}
\bibitem[{erh()}]{erhard2003c}
\bibinfo{note}{The comparison uses an estimation of the two-body loss rate
  based on the relation $\dot{N}=(4\pi\hbar a_F)/m \times N \bar{n}$ derived
  from the mean field term in the Gross-Pitaevskii equation ($a_F$ is the
  imaginary part of the scattering length at the Feshbach resonance).}
\end{thebibliography}

\end{document}